\documentclass[review]{elsarticle}

\makeatletter
\def\ps@pprintTitle{%
 \let\@oddhead\@empty
 \let\@evenhead\@empty
 \def\@oddfoot{\footnotesize\itshape
 \@journal\hfill\today
 }%
\let\@evenfoot\@oddfoot}
\makeatother

\usepackage{balance}  

\usepackage{amsmath}

\usepackage{txfonts}
\usepackage{mathptmx}
\usepackage{hyperref}
\usepackage{color}
\usepackage{booktabs}
\usepackage{textcomp}
\usepackage{microtype} 
\usepackage{ccicons}  

\usepackage{paralist}
\usepackage{url}

\usepackage[table,xcdraw]{xcolor}
\usepackage{multirow}
\usepackage{hhline}
\usepackage{graphics}
\usepackage{makecell}

\usepackage{tikz}
\usepackage{pgfplots}

\usepackage{graphicx}
\usepackage{caption}
\usepackage{subcaption}

\usepackage[flushleft]{threeparttable}

\usepackage{enumitem}

\newcolumntype{L}[1]{>{\raggedright\let\newline\\\arraybackslash\hspace{0pt}}m{#1}}
\newcolumntype{C}[1]{>{\centering\let\newline\\\arraybackslash\hspace{0pt}}m{#1}}
\newcolumntype{R}[1]{>{\raggedleft\let\newline\\\arraybackslash\hspace{0pt}}m{#1}}

\definecolor{sigP}{HTML}{C995FF}
\definecolor{sigPostHoc}{HTML}{FCDF9A}

\definecolor{ObjsColor}{HTML}{ABDDA4}
\definecolor{ImColor}{HTML}{2B83BA}
\definecolor{WordsColor}{HTML}{FFFFBF}

\makeatletter
\def\@author#1{\g@addto@macro\elsauthors{\normalsize%
    \def\baselinestretch{1}%
    \upshape\authorsep#1\unskip\textsuperscript{%
      \ifx\@fnmark\@empty\else\unskip\sep\@fnmark\let\sep=,\fi
      \ifx\@corref\@empty\else\unskip\sep\@corref\let\sep=,\fi
      }%
    \def\authorsep{\unskip,\space}%
    \global\let\@fnmark\@empty
    \global\let\@corref\@empty  
    \global\let\sep\@empty}%
    \@eadauthor={#1}
}
\makeatother

\begin{document}

\begin{frontmatter}

\title{An Exploration of Graphical Password \\ Authentication for Children}

\author{Hala Assal~\corref{cor1}}
\ead{HalaAssal@scs.carleton.ca}
\cortext[cor1]{Corresponding author}

\author{Ahsan Imran}
\ead{AhsanImran@cmail.carleton.ca}

\author{Sonia Chiasson}
\ead{chiasson@scs.carleton.ca}

\address{School of Computer Science, Carleton University\\1125 Colonel By Dr., Ottawa, ON Canada}
\fntext[myfootnote]{This project was supported by a grant from the Canadian Internet Registration Authority (CIRA)'s Community Investment Program.
}

\begin{abstract}
In this paper, we explore graphical passwords as a child-friendly alternative for user authentication.  We evaluate the usability of three variants of the PassTiles graphical password scheme for children, and explore the similarities and differences in performance and preferences between children and adults while using these schemes. Children were most successful at recalling passwords containing images of distinct objects. Both children and adults prefer graphical passwords to their existing schemes, but password memorization strategies differ considerably between the two groups. Based on our findings, we provide recommendations for designing more child-friendly authentication schemes.
\end{abstract}


\end{frontmatter}



\section{Introduction}\label{sec:intro}

Children constitute a significant portion of Internet users; 99\% of Canadian children aged 8 to 15 years are active online~\cite{CANusrs} and the statistics are similar in other countries~\cite{holloway2013zero}. 
 Despite abundant authentication research, the literature on user authentication specifically for children is relatively sparse. In fact, the main focus of online security for children has been on designing methods for adults to protect their children~\cite{channakeshava2008providing,colella2010system}, or to educate children about the dangers of Internet usage~\cite{Rode:2009:DPD:1671011.1671041}.

 Children do not necessarily have the same skills and preferences as adults, thus a child-oriented authentication system that better suits children's privacy and security needs is long overdue. In this paper, we seek to address the gap by exploring the usability of graphical passwords for children, and provide recommendations for developing authentication mechanisms for children that are more adaptable to their needs. We conducted two user studies with both children and adults in which they interacted with three graphical variants of the PassTiles~\cite{stobert2012visual} password scheme: Objects, Image, and Words PassTiles. We explore the similarities and differences in children's and adults' performance and perception. We found that \emph{both} children and adults prefer graphical passwords to their existing schemes. This paper has the following contributions.

 \begin{enumerate}
 \item Attempts to address the gap of child-oriented user authentication by exploring the usability of graphical passwords for children.
 \item Provides empirical results comparing children's and adults' performance and preferences while using three different graphical password schemes.
 \item Provides recommendations for adapting user authentication for children.
 \end{enumerate} 
 
\paragraph{Organization} Section~\ref{sec:background} presents a background on user authentication.  Section~\ref{sec:Passtiles} presents the original PassTiles graphical password scheme and our adaptation. We present our user studies in Section~\ref{sec:methodology}, and the results in Section~\ref{sec:results}.  The implications of the results and general recommendations for children's authentication are discussed in Section~\ref{sec:discussion}.

\section{Background}\label{sec:background}

\subsection{Authentication}

The most commonly used authentication schemes are knowledge-based~\cite{SecurityBook,Schneier:2004:SA:971564.971595,6234436}, where the user memorizes a shared secret such as a password. Research has demonstrated that text-based passwords suffer considerabily from \textit{both} security and usability issues \cite{Biddle:2012:GPL:2333112.2333114,6234436}. %
Bonneau~\textit{et al.}~\cite{6234436} developed a framework for evaluating alternative authentication schemes based on usability, deployability, and security benefits. They found that no known scheme is ultimately better, suggesting that when choosing an alternative to text passwords, one needs to balance the advantages and tradeoffs, and choose the scheme that mostly fits one's needs.

Despite abundant authentication research, the literature on user authentication specifically for children is relatively sparse. We study graphical passwords~\cite{Biddle:2012:GPL:2333112.2333114} as an alternative authentication scheme for children. Psychology studies have acknowledged the human brain's superiority in recognizing and recalling visual information in contrast to textual information~\cite{kirkpatrick1894experimental,Biddle:2012:GPL:2333112.2333114}. Graphical passwords, a form of knowledge-based authentication, aim to utilize this human feature to reduce the user's cognitive load of memorizing passwords~\cite{Chiasson2007}. Graphical passwords can be categorized according to the cognitive task necessary to remember the password: recognition, recall, and cued recall~\cite{Raaijmakers1992}. Recognition is the least cognitively burdensome, where a user needs to decide whether the information presented to her matches what she had already memorized. Recall, on the the other end of the difficulty spectrum, requires the user to remember the information memorized without any clues. Cued recall offers some cues to trigger the user's stored memory~\cite{Biddle:2012:GPL:2333112.2333114,Chiasson2007}.

We focus our study on two recognition based graphical passwords (Objects and Words PassTiles) and one cued recall graphical password (Image PassTiles)~\cite{stobert2012visual}.

\subsection{Children and Passwords}

Recently researchers have taken interest in understanding how children perceive passwords to develop security guidelines and recommendations for children. Lorenz~\textit{et al.}~\cite{Lorenz2013} found that younger children were more receptive to security advice than teenagers and adults. Studies by Read~\textit{et al.}~\cite{Read:2012:DTP:2307096.2307125} and Coggins~\cite{doi:10.1080/07380569.2013.807719} investigated children's knowledge of text passwords. Both studies found that children, without any formal training, have at least some understanding of the purpose of passwords and how to create strong ones. On the other hand, Zhang-Kennedy~\textit{et. al.}~\cite{zhang-kennedy2016-kidsmobile-idc} found that adults (parents and teachers) are primarily responsible for maintaining children's online credentials and the majority of children do not understand the reason why passwords should remain secret. Coggins recommends integrating the concept of computer passwords in elementary school curriculum, to teach children early on best practices of computer safety. He also suggests training children to memorize strings of random characters. Read~\textit{et al.}~\cite{Read:2009:UMP:1671011.1671046} recommend using analogies when explaining security concepts to children, e.g., keeping their game consoles safe in a drawer as an analogy to explain the importance of keeping their passwords safe.

\subsection{Authentication for Children}

Two US patents have been filed for relevant authentication schemes. Each took different approaches to authentication for children. Colella~\cite{colella2010system} proposed an authentication system that relies primarily on biometric identification. Using this authentication system, a third party Application Service Provider (ASP) assigns each child a ``Safe Card" fingerprint scanner. The ASP maintains a database of approved websites for each child. To login, the child connects her Safe Card to her computer, performs local authentication on it, and the card in turn opens the ASP's webpage and authenticates the child to the ASP. The child can then access any of the pre-approved websites through the ASP's webpage. M. Renaud and Mulji's~\cite{renaud2007authentication}  proposed ``Little Bo Peep" scheme guides users to create their own versions of familiar age-appropriate stories (e.g., fairy tales), which are then used as their authentication tokens. They claim that the personalization renders the stories more memorable than traditional authentication schemes.

On the adacemic side, K. Renaud~\cite{5341710} proposed an authentication system comprised of user-drawn icon-sized images (\emph{Mikons}~\cite{Mikons}) with password space equivalent to $15.9\ bits$. Renaud suggests its use for low-risk systems used by children. 
It was tested by children (11 and 12 years old) to access their online homework system. To sign up, each student drew four Mikons to represent their password, then the class teacher approved their choice of Mikons. To login, students choose their drawings from other distractor Mikons. They were presented with four sets of challenges, each containing 15 distractors and one of their Mikons. Over the period of 8 months, students logged into their homework system once every 2 months. Overall the system recorded an 87\% success rate. 
Mendori~\textit{et al.}~\cite{1186069} designed a password system for grade 1 Japanese students, unfamiliar with the Roman alphabet and thus were unable to use traditional text passwords. Mendori's interface displays preset icons and symbols on a grid, and the child clicks on the ones that comprise their password. The researchers experimented with three different configurations, all with password space equivalent to $11.9\ bits$, however little has been published about the system's usability.


\section{PassTiles}\label{sec:Passtiles}
PassTiles~\cite{stobert2012visual} is a graphical password system designed to facilitate the comparison of different memory retrieval types by offering several configurable parameters. It presents the user with a grid of tiles, and the password is composed of a subset of these tiles. The passwords are system assigned and generated through the Multiple Versatile Passwords (MVP) framework~\cite{Chiasson2012}.

During the password memorization phase, the \emph{password tiles} are visually highlighted with a coloured border. PassTiles users practice entering their password by clicking on the highlighted tiles in any order. When the user clicks a highlighted tile, the highlight colour changes to indicate selection. A visual, and numeric, counter is decremented with every click to show the number of tiles yet to be clicked.  During login, the grid appears without highlighting, but the counter is displayed. To login, users click on their memorized tiles in no particular order.  

We used three PassTiles schemes: Image, Objects, and Words. Each has 48 square tiles in a 6$\times$8 grid, and each password is composed of 5 tiles. Image PassTiles~\cite{stobert2012visual} leverages cued recall by superimposing the tiles on a user-chosen background image,\footnote{The background image is user-chosen from a system-provided set.} which users can use as a cue to recall the location of their password tiles. Objects~\cite{stobert2012visual}, and Words~\cite{Wright:2012:YSY:2335356.2335367} PassTiles leverage recognition. Each tile in Objects PassTiles contains an image of an object, and the password is composed of a set of these objects. With each login, the tiles are shuffled, and users must rely on recognition to remember their password objects, rather than recalling the tile positions.  Words PassTiles is similar to Objects, except it displays short simple words on the tiles. 

\begin{figure}
    \centering
    \begin{subfigure}[b]{0.33\textwidth}
        \frame{\includegraphics[width=\textwidth]{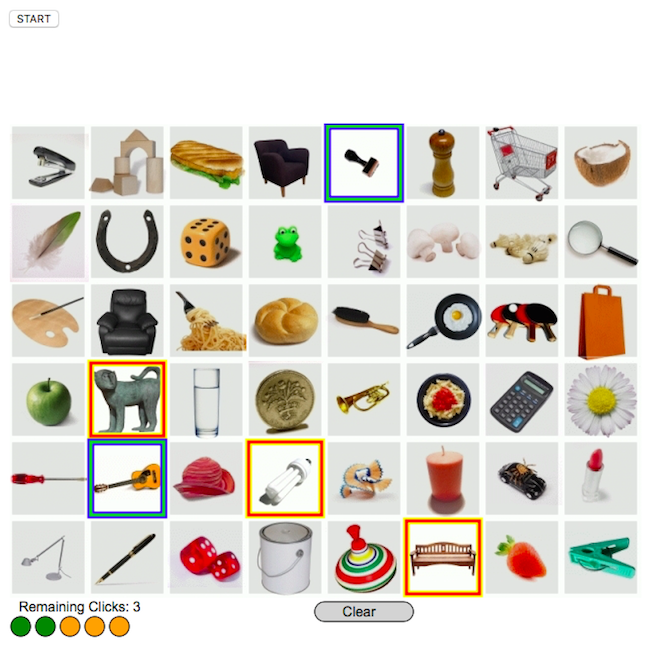}}
        \caption{Objects PassTiles}
        \label{fig:objectsMem}
    \end{subfigure}%
    ~
    \begin{subfigure}[b]{0.33\textwidth}
        \frame{\includegraphics[width=\textwidth]{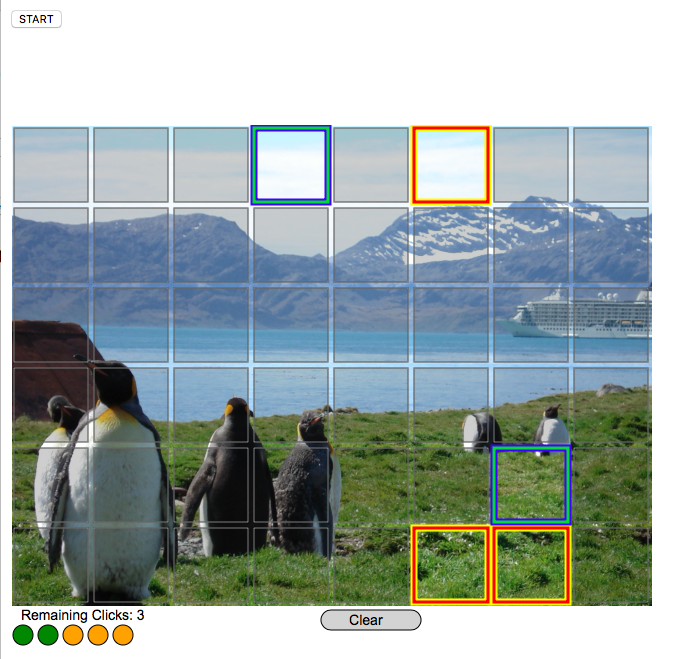}}
        \caption{Image PassTiles}
        \label{fig:imageMem}
    \end{subfigure}%
    ~
    \begin{subfigure}[b]{0.33\textwidth}
        \frame{\includegraphics[width=\textwidth]{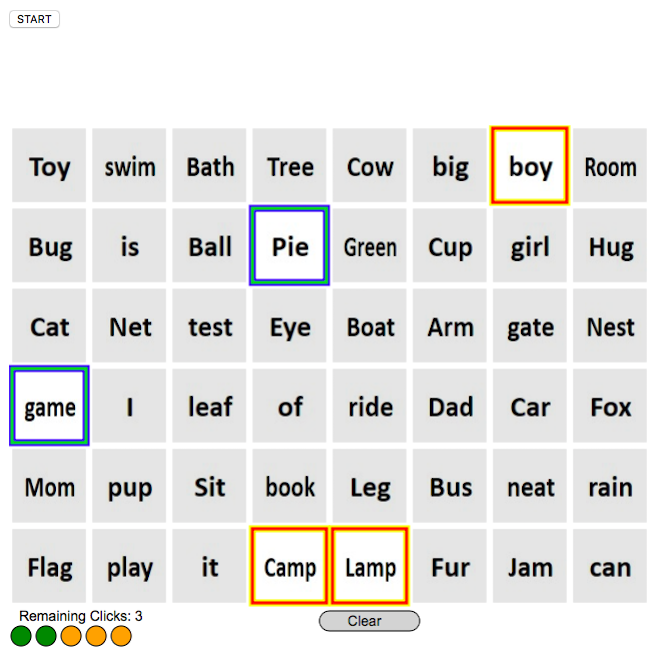}}
        \caption{Words PassTiles}
        \label{fig:wordsMem}
    \end{subfigure}
    \caption{Adapted PassTiles interface during memorization phase with highlighted password tiles. The orange highlight indicates the tile is yet to be clicked. A counter at the bottom left  indicates the remaining number of tiles to be clicked (3 in this example).}\label{fig:schemesMem}
\end{figure}

\begin{figure}
    \centering
    \begin{subfigure}[b]{0.33\textwidth}
        \includegraphics[width=\textwidth]{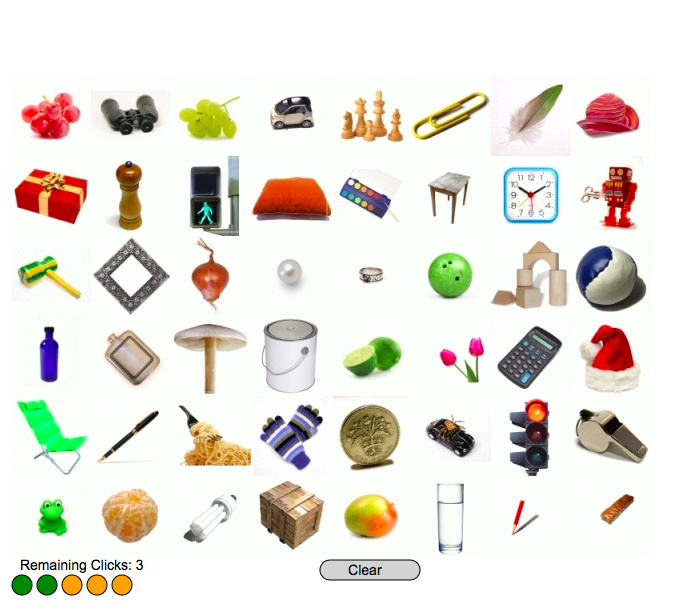}
        \caption{Objects PassTiles}
        \label{fig:objectsLogin}
    \end{subfigure}%
    \begin{subfigure}[b]{0.33\textwidth}
        \includegraphics[width=\textwidth]{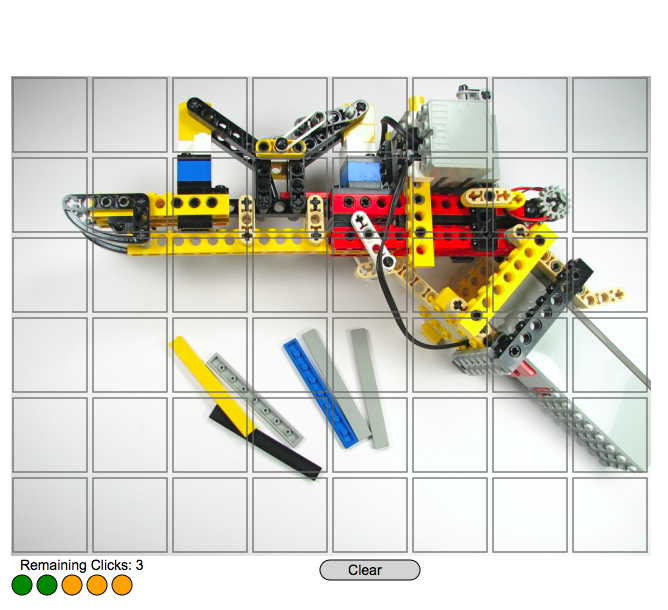}
        \caption{Image PassTiles}
        \label{fig:imageLogin}
    \end{subfigure}%
    \begin{subfigure}[b]{0.33\textwidth}
        \includegraphics[width=\textwidth]{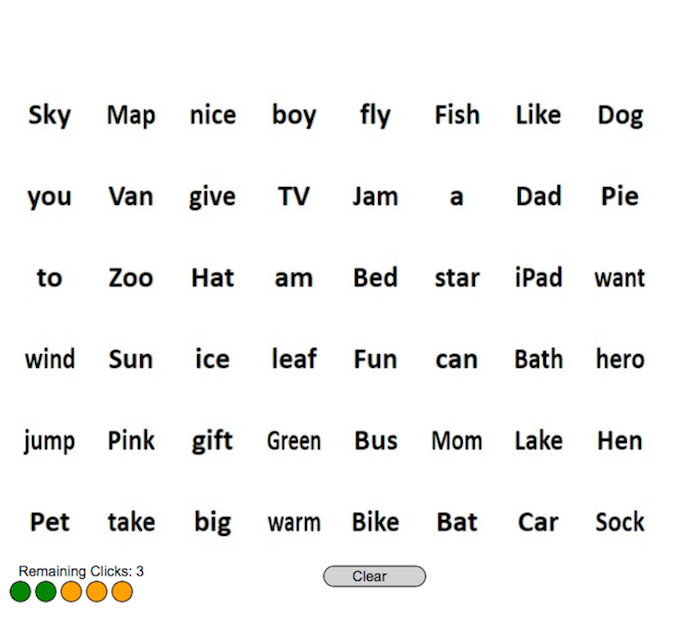}
        \caption{Words PassTiles}
        \label{fig:wordsLogin}
    \end{subfigure}
    \caption{Adapted PassTiles interface during login, where clicked tiles are not highlighted, but the counter is decremented.}\label{fig:schemesLogin}
\end{figure}

 For the purpose of our study, we simplified the original PassTiles interface by removing non-essential features and simplifying the language to be better suited for children. The existing image set on MVP was filtered and more age-appropriate images were added. For Words PassTiles, the existing set of words was modified to shorter and simpler ones. Figure~\ref{fig:schemesMem} shows the three schemes during the memorization phase where 2 out of the 5 password tiles have been selected. The counter, at the bottom left of the screen, also shows that 3 tiles remain. Figure~\ref{fig:schemesLogin} shows the three schemes during the login phase. The interface does not highlight clicked tiles; in this example, the counter indicates that user has clicked two tiles.

\subsection{Security of PassTiles}
Knowledge-based authentication systems must be resilient against two main types of attacks: capture and guessing attacks~\cite{Biddle:2012:GPL:2333112.2333114}. Capture attacks involve acquiring the actual password or part thereof (e.g., using keyloggers or observing password entry through shoulder-surfing). A guessing attack is when the attacker tries to guess the password. Depending on circumstances, an attacker may be able to exhaustively search the theoretical password space (the total set of passwords an authentication scheme allows), guaranteeing success, or may be limited in the number of guesses. Using knowledge of password distributions, an attacker can prioritize higher probability passwords (ones that users are more likely to use). Thus, to be resilient to guessing attacks, authentication systems should have an appropriately large theoretical password space and prevent predictable patterns in user-chosen passwords~\cite{Biddle:2012:GPL:2333112.2333114}.  

Recall, a PassTiles password is composed of 5 tiles on a grid of 6$\times$8 tiles, and the order of password entry is insignificant. The size of the theoretical password space for PassTiles is  $\log_2 \scriptsize \dbinom{\# rows \times \# columns}{password\ length} \normalsize= \log_2 \scriptsize \dbinom{6\times8}{5}\normalsize = 21\ bits$. Flor{\^e}ncio~\textit{et al.}~\cite{186500} suggest that a password space of 20 bits is sufficient to protect against online guessing attacks. 
To reduce the success of guessing attacks, PassTiles uses system-generated passwords to utilize the full theoretical password space. In addition, system-assigned passwords protect against targeted guessing attacks, as users do not choose personally relevant elements. PassTiles does not highlight clicked tiles during password entry to reduce the risk of shoulder-surfing attacks.


\section{Methodology}\label{sec:methodology}

We conducted two user studies, one with children and one with adults. The adult study serves as our control condition. 
Our studies used a within subject design; each participant tested all three schemes (Objects, Image, and Words). For both studies, we used a balanced presentation order for the password schemes to minimize training effects. Both studies were reviewed and cleared by our institution's Research Ethics Board.

\subsection{Procedure}
\paragraph{\textbf{Children Study}} Each study session consisted of 5 phases, and lasted an average of 25 minutes. Each participant received a \$10 gift card as compensation, even if they withdrew from the study.

\subsubsection*{Phase 1: Introduction and Consent}

\begin{description}
\item [Consent]   The participants' parents signed a consent form agreeing to have their children participate in the study and be audio-recorded. The Children subsequently provided oral consent. 
\item [Introduction]  We explained to the participants the tasks they would be performing. We showed them printed snapshots of the three password schemes and explained verbally how the system worked and how they would enter their passwords once they were memorized.
\end{description}

\subsubsection*{Phase 2: Memorization and First login}
Phase 2 was repeated three times, once for each password scheme. Participants completed Phase 2a followed by 2b for one scheme before moving to the next. They were given the chance to login only once for each scheme.

\begin{description}
\item [Phase 2(a): Memorize Password] \hfill \\  Participants memorized the system-assigned password presented on the screen.
 Participants could practice entering their password as many times as they wanted. Once the participant had memorized the password, she was asked to click the start button.
\item [Phase 2(b): First Login] \hfill \\ On clicking the start button, participants were presented with the PassTiles interface with no highlighted tiles.  Participants clicked on the tiles they thought composed their password, and the counter below the grid was incremented with every click. Once they had made five clicks, a popup box informed them if it was correct. 
\end{description}

\subsubsection*{Phase 3: Interview}
We interviewed participants to gather some insight about their preferences and perceptions of the three password schemes. The interview phase, which lasted for approximately 12 minutes, also served as a distraction period between the first and second login attempts (described in Phase 4). All interviews were audio recorded.

\subsubsection*{Phase 4: Second Login}
To test memorability, participants were asked to perform a second attempt. The three schemes were presented to each participant in the same order as in the first login attempt, and each participant had only one attempt per scheme.

\paragraph{\textbf{Adult Study}} It followed the same procedure, with the following exceptions. The consent form was signed by the participants themselves, each participant received \$10 compensation in cash, and each session lasted an average of 18 minutes.

\subsection{Environment and Equipment}
The children sessions took place either at our research labs or at a public library. Both offered a quiet, low-distraction environment. The adults sessions took place in our lab.

We used a Sony VAIO laptop with a touch-screen running Windows~8. Participants were instructed to use the mouse when interacting with PassTiles; however, three children found the mouse too difficult to use and switched to using the touchscreen. 

\subsection{Participant Demographics}
\subsubsection{Children Study}
We recruited 25 children between 7 and 12 years of age ($mean=9.5$ years), ten boys and fifteen girls. All children were accompanied by a parent. Twenty-four participants had previously used at least one type of authentication on their tablets. Table~\ref{table:demogChildren} shows the distribution of children's experience with existing authentication schemes. Twenty participants reported their parents taught them how to create passwords, two were taught by their older siblings, and two were self-taught.

\renewcommand{\arraystretch}{0.7}
\begin{table}[]
\centering
\caption{Distribution of authentication schemes children have used on their tablets.}
\label{table:demogChildren}
\begin{tabular}{rccccc}
\toprule
\multirow{2}{*}{Authentication scheme} & \multirow{2}{*}{None} & \multirow{2}{*}{PINs} & \multicolumn{2}{c}{Passwords} & \multirow{2}{*}{Pattern} \\ \cmidrule(lr){4-5}
                                       					&                       	&                       & alphanumeric  & letters only  &                       \\ \bottomrule
Number of children                          		&1                       		&10            &9               		&6               				  &2 \\ \bottomrule
\end{tabular}
\end{table}

\subsubsection{Adult Study}
We recruited 25 adults for this study. Twenty participants were between 18 and 30 years of age and five were over 30. Fourteen participants were male and eleven female. Five participants had a high school diploma, twelve had a Bachelor's degree, six had a Master's degree and two had Doctoral degrees. All participants were familiar with the use of passwords and had experience using multiple passwords for different online accounts.


\section{Results}\label{sec:results}

We explored user performance through time and login success measures, and user preferences through interview questions.

We conducted Shapiro-Wilk test to check the normality of children's and adults' memorization and login times (first and second attempt) for the three schemes. The distribution of times significantly deviated from the normal distribution ($p<0.05$), thus for these variables, we use non-parametric statistical tests.

For the following statistical tests, unless otherwise stated, we conduct Friedman tests to determine the differences between the three authentication schemes. In case of significant difference, we follow up using Bonferroni-corrected Wilcoxon post hoc tests. For these tests the calculated degrees of freedom is 2.

\subsection{Memorization Time}\label{sec:MemTime}

\begin{figure}
\centering
\includegraphics[scale=0.35]{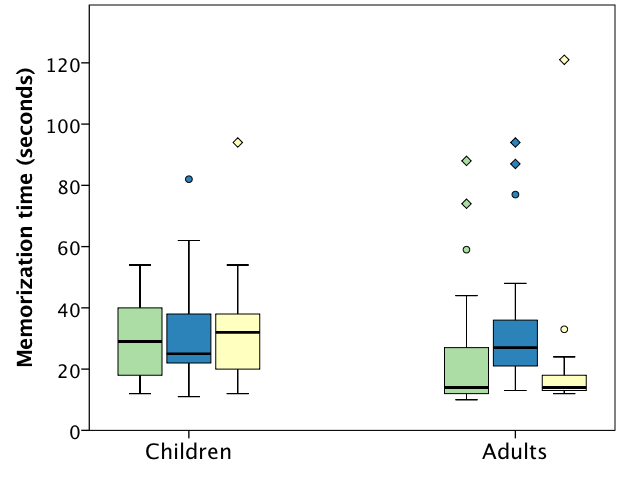}
\caption{Memorization time for \colorbox{ObjsColor}{Objects}, \colorbox{ImColor}{Image}, and \colorbox{WordsColor}{Words}}
\label{fig:memTime}
\end{figure}

We recorded the time spent memorizing each password, from the time the assigned password appeared until the participant clicked the start button to perform the first login attempt. Figure~\ref{fig:memTime} shows the descriptive statistics for the time spent by children and adults memorizing each scheme. Table~\ref{table:timesStats} presents results of the statistical tests comparing times per scheme for the children and adults separately. Average times per scheme (in seconds) range from 29.5 to 32.6 for children and 20.1 to 33.5 for adults, with several outliers. 

We found no statistical difference between schemes in children's memorization time. However, there was a statistical difference for adults; the post hoc test indicated that adults took significantly longer memorizing their Image password than the other two schemes. Since memorization time is a reflection of how much effort participants chose to devote to memorizing their password, we conclude that children did not perceive one scheme to be harder to memorize than the other, while adults perceived Image to be more difficult.

\subsection{Login Times}
\begin{figure}
    \centering
    \begin{subfigure}[b]{0.49\textwidth}
        \includegraphics[width=\textwidth]{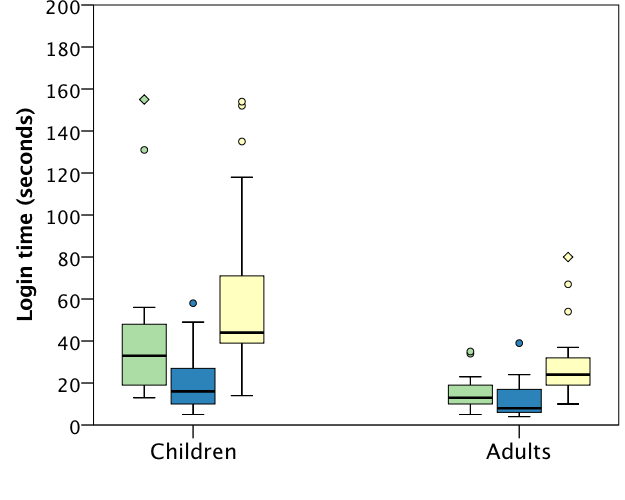}
        \caption{First attempt}
        \label{fig:loginFirstBox}
    \end{subfigure}
    ~ %
    \begin{subfigure}[b]{0.49\textwidth}
        \includegraphics[width=\textwidth]{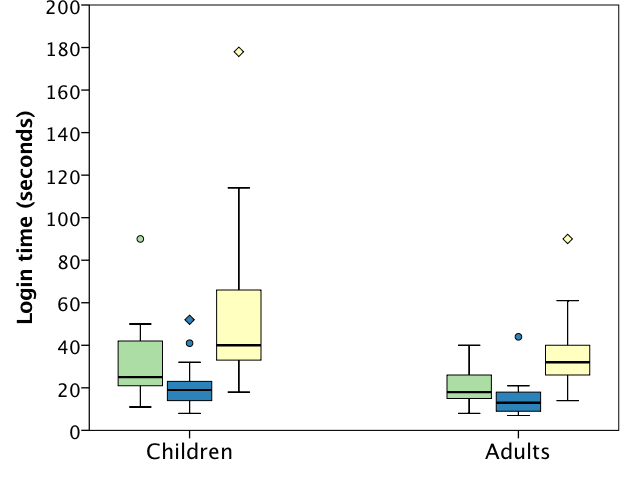}
        \caption{Second attempt}
        \label{fig:loginSecondBox}
    \end{subfigure}
    \caption{Time (\emph{in second}) to login by children and by adults using \colorbox{ObjsColor}{Objects}, \colorbox{ImColor}{Image}, and \colorbox{WordsColor}{Words}}
    \label{fig:loginBox}
\end{figure}

The time spent entering each password was recorded, from the time participants clicked the start button until they clicked 5 tiles. Figure~\ref{fig:loginBox} shows the descriptive statistics for the time participants spent logging in using the three schemes. Table~\ref{table:timesStats} presents results of the statistical tests measuring the effect of scheme for the first and second login attempts respectively, per study. Average login times (in seconds) range from 20.5 to 60.6 for children and 11.7 to 35.7 for adults, with much less variation for adults.

\subsubsection{First Attempt}
On the first attempt, children were significantly faster logging in using Image compared to the two other schemes. Children were also significantly faster logging in using Objects compared to Words.  Similarly, login times for adults' first attempt were significantly different for the three schemes. The post hoc test shows that the login time for Words was significantly slower than that of Objects and Image. We conclude that on the first login attempt, children found Image the easiest to recall, followed by Objects, and both children and adults found Words the hardest scheme to recall.

\subsubsection{Second Attempt}
We found a statistical difference between children's second login times; the post hoc test shows that the login time for Words was statistically slower than that of Objects and Image. Additionally, adults' second attempt login times for Words was statistically slower than Objects, which in turn was statistically slower than Image. We conclude that, on second attempt, adults found Image the easiest to recall, followed by Objects, and both children and adults found Words the hardest scheme to recall.

\renewcommand{\arraystretch}{0.5}
\begin{table}[]
\centering
\caption{Results from Friedman test and Bonferroni-corrected Wilcoxon post hoc test looking at differences between the three schemes in memorization and login times by children and by adults.}
\label{table:timesStats}
\scalebox{0.85}{
\begin{tabular}{R{2cm}C{1cm}C{0.7cm}C{0.7cm}C{0.8cm}C{0.8cm}C{1cm}C{0.7cm}C{0.7cm}C{0.8cm}C{0.8cm}}
\toprule
         & \multicolumn{5}{c}{Children}                                             & \multicolumn{5}{c}{Adults}                \\[-2pt] \cmidrule(l){2-6}\cmidrule(l){7-11}    
         & \multicolumn{2}{c}{Friedman test} & \multicolumn{3}{c}{Post hoc}        & \multicolumn{2}{c}{Friedman test} & \multicolumn{3}{c}{Post hoc} \\ \cmidrule(l){2-3}\cmidrule(l){4-6} \cmidrule(l){7-8}\cmidrule(l){9-11}  %
	time 			& $X^2$	& $p$		& O-I	& O-W		& I-W 		& $X^2$	& $p$		& O-I      & O-W     & I-W     \\ \bottomrule %
						&			&			&		&			&			&			&			&		 &		& \\[-8pt] %
	Memorization  		&1.354   	&.508       	&     	&     		&     		&17.111      &\cellcolor{sigP}.000		&\cellcolor{sigPostHoc}.006     &1.000    &\cellcolor{sigPostHoc}.000 		\\
     1$^{st}$ login    		&30.081	&\cellcolor{sigP}.000		&\cellcolor{sigPostHoc}.007	&\cellcolor{sigPostHoc}.049		&\cellcolor{sigPostHoc}.000       	&22.566     &\cellcolor{sigP}.000		&.537       &\cellcolor{sigPostHoc}.003     	  &\cellcolor{sigPostHoc}.000  \\
     2$^{nd}$ login 		&29.714    &\cellcolor{sigP}.000        	&.071    &\cellcolor{sigPostHoc}.006   	 &\cellcolor{sigPostHoc}.000          &26.929     &\cellcolor{sigP}.000		&\cellcolor{sigPostHoc}.022         &\cellcolor{sigPostHoc}.040    		&\cellcolor{sigPostHoc}.000	 \\
         \bottomrule
\end{tabular}
}%
    \vspace{-0.2cm}
    \begin{tablenotes}
    \scriptsize \item O: Objects, I: Image, and W: Words. Colour indicates statistically significant results 
    \end{tablenotes}
\end{table}

\subsection{Login Success}

\begin{figure}
    \centering
    \begin{subfigure}[]{0.49\textwidth}
        \includegraphics[width=\textwidth]{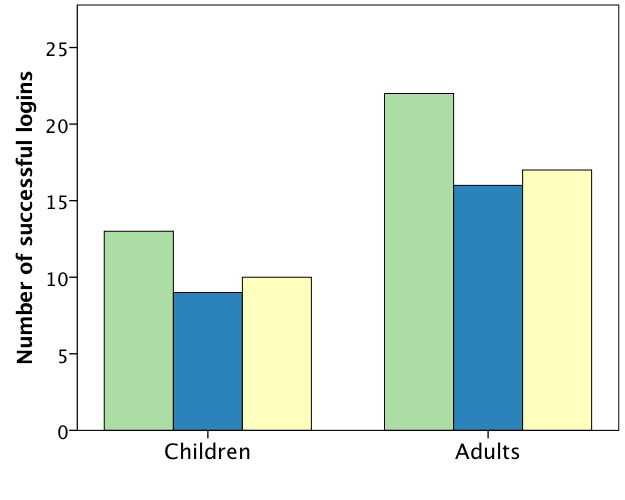}
        \caption{First attempt}
        \label{fig:SuccFailFirst_Desc}
    \end{subfigure}
    ~ %
    \begin{subfigure}[]{0.49\textwidth}
        \includegraphics[width=\textwidth]{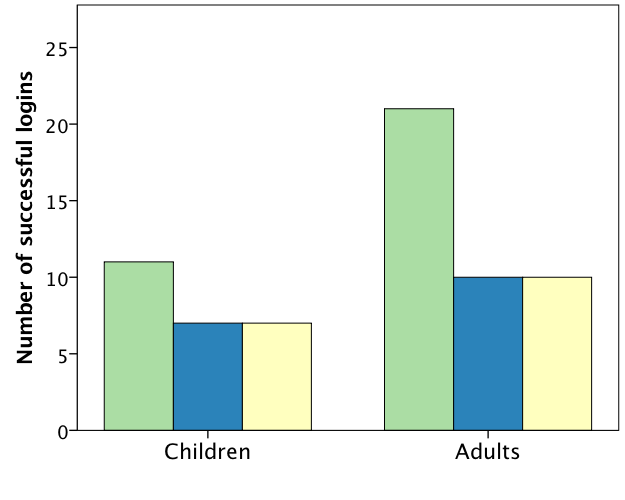}
        \caption{Second attempt}
        \label{fig:SuccFailSecond_Desc}
    \end{subfigure}
    \caption{Successful login attempts by children and by adults using \colorbox{ObjsColor}{Objects}, \colorbox{ImColor}{Image}, and \colorbox{WordsColor}{Words}}
    \label{fig:SuccFail_Desc}
\end{figure}

We recorded whether each login was a successful or a failed attempt. Figure~\ref{fig:SuccFail_Desc} shows the number of participants' successful logins. To test if there is a significant effect of scheme in each study, we use Cochran's Q test. We follow up on statistically significant results by conducting Bonferroni-corrected McNemar post hoc tests. The results of the statistical tests is presented in Table~\ref{table:loginSuccessStats}.

\subsubsection{First Attempt}
On first attempt, we found no effect of scheme on login success for children or adults. 

\subsubsection{Second Attempt}
There was no effect of scheme on login success for children's second attempt. However, we found statistical difference between the three schemes for adults. Adults were significantly more successful logging in using Objects than the other two schemes. %

We built a logistic regression model to determine the effect of the length of memorization time on the likelihood that participants successfully logged in on the second attempt. However, the model was not significant, for neither children nor adults, thus memorization time cannot be used as a predictor for the outcome of the login attempt.

\renewcommand{\arraystretch}{0.5}
\begin{table}[]
\centering
\caption{Results from Cochran's Q test and Bonferroni-corrected McNemar post hoc test looking at difference in login success between the three schemes for children and for adults. 
The calculated degrees of freedom is 2.}
\label{table:loginSuccessStats}
\scalebox{0.85}{
\begin{tabular}{R{2cm}C{1cm}C{0.7cm}C{0.7cm}C{0.8cm}C{0.8cm}C{1cm}C{0.7cm}C{0.7cm}C{0.8cm}C{0.8cm}}
\toprule
         & \multicolumn{5}{c}{Children}                                             & \multicolumn{5}{c}{Adults}                \\[-2pt] \cmidrule(l){2-6}\cmidrule(l){7-11}    
         & \multicolumn{2}{c}{Cochran's Q test} & \multicolumn{3}{c}{Post hoc}        & \multicolumn{2}{c}{Cochran's Q test} & \multicolumn{3}{c}{Post hoc} \\ \cmidrule(l){2-3}\cmidrule(l){4-6} \cmidrule(l){7-8}\cmidrule(l){9-11}  %
	login success 			& $X^2$	& $p$		& O-I	& O-W		& I-W 		& $X^2$	& $p$		& O-I      & O-W     & I-W     \\ \bottomrule %
     1$^{st}$ attempt    		&2			&.368		&		&			&        		&4.133     	&.127		&	        &     	    &  \\[4pt]
     2$^{nd}$ attempt 		&2.909    	&.234         &   		&   	 		&          		&12.737     &\cellcolor{sigP}.002		&\cellcolor{sigPostHoc}.006         &\cellcolor{sigPostHoc}.006   &1.000		 \\%
         \bottomrule
\end{tabular}
}%
    \vspace{-0.2cm}
    \begin{tablenotes}
    \scriptsize \item O: Objects, I: Image, and W: Words. Colour indicates statistically significant results
    \end{tablenotes}
\end{table}

\subsection{Degree of correctness}
Compared to previous literature, adults' success rate is considerably lower. While no comparison is available for children, they also struggled with login. This could have been influenced by our study design, stressing memory by having three passwords memorized in a short time and allowing only one login attempt at a time. To explore further the reasons for low login success, we examined partially correct responses. We counted the number of correct tiles for each attempt. Figure~\ref{fig:correctLogins} shows the descriptive statistics for the number of correctly chosen tiles.

\subsubsection{First Attempt}
The median for each scheme is 4 or 5 correct tiles out of 5 on the first attempt, suggesting that participants had reasonable success at remembering most of their password. We found no effect of scheme on the degree of correctness for children, nor for adults. 

\subsubsection{Second Attempt}

\begin{figure}
    \centering
    \begin{subfigure}[b]{0.49\textwidth}
        \includegraphics[width=\textwidth]{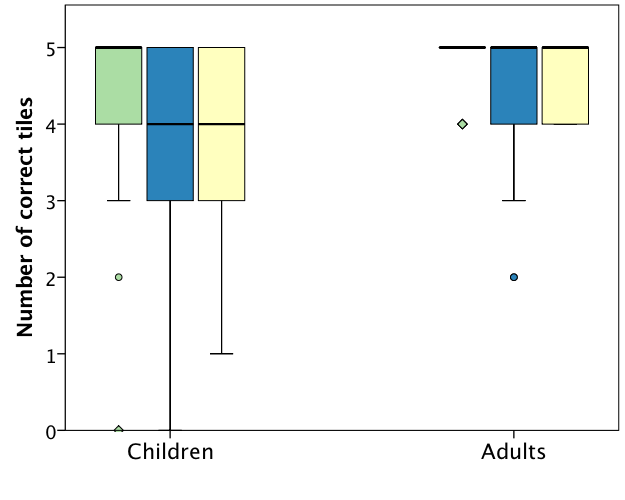}
        \caption{First login attempt}
        \label{fig:correctFirst}
    \end{subfigure}
    ~ %
    \begin{subfigure}[b]{0.49\textwidth}
        \includegraphics[width=\textwidth]{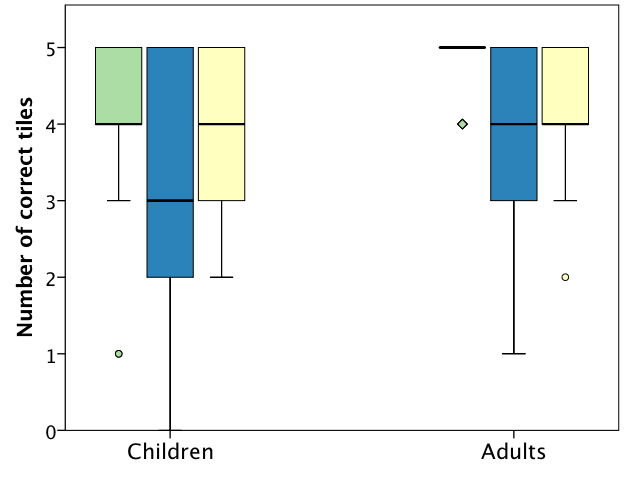}
        \caption{Second login attempt}
        \label{fig:correctSecond}
    \end{subfigure}
    \caption{Number of correct tiles selected on each login attempt using \colorbox{ObjsColor}{Objects}, \colorbox{ImColor}{Image}, and \colorbox{WordsColor}{Words}}
    \label{fig:correctLogins}
\end{figure}

The degree of correctness for the second attempt showed more variability, particularly for the children. A Friedman test on the degree of correctness of children's second login attempt shows a statistical difference between the schemes, however, the pairwise comparison cannot identify with significant confidence which pairs differ. Looking at the data (see Figure~\ref{fig:correctSecond}), we suspect that with a larger sample size, Objects would have the highest level of correctness while Image would be the worst. On the other hand, the degree of correctness for adults' second login attempt showed a statistical difference, and the post hoc test confirmed that adults' degree of correctness for Objects was significantly higher than that of Image and Words. We conclude that on second login attempt, adults found Objects the easiest to correctly recall, and children appear to be following the same trend.

\renewcommand{\arraystretch}{0.5}
\begin{table}[]
\centering
\caption{Results from Friedman test and Bonferroni-corrected Wilcoxon post hoc test looking at differences between the three schemes in the number of correct tiles chosen by children and by adults.}
\label{table:correctnessStats}
\scalebox{0.85}{
\begin{tabular}{R{2cm}C{1cm}C{0.7cm}C{0.7cm}C{0.8cm}C{0.8cm}C{1cm}C{0.7cm}C{0.7cm}C{0.8cm}C{0.8cm}}
\toprule
         & \multicolumn{5}{c}{Children}                                             & \multicolumn{5}{c}{Adults}                \\[-2pt] \cmidrule(l){2-6}\cmidrule(l){7-11}    
         & \multicolumn{2}{c}{Friedman test} & \multicolumn{3}{c}{Post hoc}        & \multicolumn{2}{c}{Friedman test} & \multicolumn{3}{c}{Post hoc} \\ \cmidrule(l){2-3}\cmidrule(l){4-6} \cmidrule(l){7-8}\cmidrule(l){9-11}  %
	correctness 			& $X^2$	& $p$		& O-I	& O-W		& I-W 		& $X^2$	& $p$		& O-I      & O-W     & I-W     \\ \bottomrule %
     1$^{st}$ attempt    		&3.077		&.215		&		&			&        		&4.978     	&.083		&	        &     	    &  \\[4pt]
     2$^{nd}$ attempt 		&6.200    	&\cellcolor{sigP}.045        &.198   		&.231   	 		&1.000          		&14.94     &\cellcolor{sigP}.001		&\cellcolor{sigPostHoc}.014        &\cellcolor{sigPostHoc}.022    &1.000		 \\%
         \bottomrule
\end{tabular}
}%
    \vspace{-0.2cm}
    \begin{tablenotes}
    \scriptsize \item O: Objects, I: Image, and W: Words
    \end{tablenotes}
\end{table}

\subsection{Interview Results}
We explored participants' perceptions and knowledge of online privacy and security, and asked them specific questions about the three schemes they tested; we asked participants:
\begin{itemize}
 \item Which type of password did you like the most? \emph{[Preferred scheme]}
 \item Which one was the most difficult? \emph{[Most Difficult scheme]}
 \item Which of the passwords do you think is the safest? \emph{[Safest scheme]}
\end{itemize}

For each question, we tabulated the number of participants that identified a scheme in their response. Figure~\ref{fig:interviewBars} summarizes participants' responses. The Safest scheme question has only 24 responses; one child could not identify the safest scheme. We conducted One-Way Chi Square tests  to determine whether a scheme stood out as the most preferred, the most difficult, the safest, by comparing to expected even distributions of 8.5 responses per scheme. 

\subsubsection{Preferred Scheme}
We did not find statistical evidence of one scheme being more preferred by children, nor by adults. Looking at the distributions in Figure~\ref{fig:interviewBars}, we see very similar patterns between children and adults; Objects was the preferred scheme for 48\% of the children and 44\% of adults, followed by Words then Image. 
  A larger sample would be needed to confirm this trend.   

\subsubsection{Most Difficult Scheme}
We found no statistical evidence of one scheme being perceived by children as the most difficult. However, adults perceived Objects to be the least difficult ($X^2(2, N=25)=7.760, p<0.05$). This coincides with it being preferred by more adults compared to Objects and Image.  

\subsubsection{Safest Scheme}
\begin{figure}
    \centering
    \begin{subfigure}[b]{0.49\textwidth}
        \includegraphics[width=\textwidth]{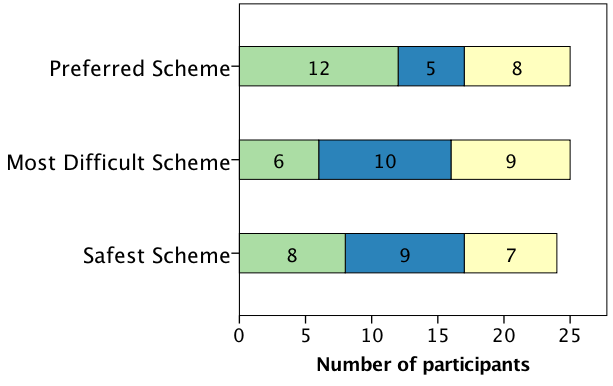}
        \caption{Children}
        \label{fig:interviewChildrenBar}
    \end{subfigure}
    ~ %
    \begin{subfigure}[b]{0.49\textwidth}
        \includegraphics[width=\textwidth]{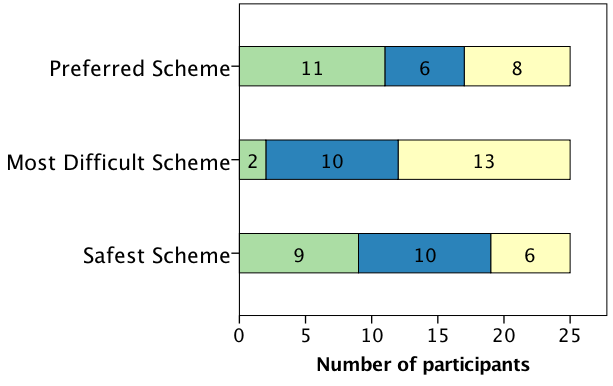}
        \caption{Adults}
        \label{fig:interviewAdultsBar}
    \end{subfigure}
    \caption{Interview Results showing the number of participants choosing \colorbox{ObjsColor}{Objects}, \colorbox{ImColor}{Image}, and \colorbox{WordsColor}{Words}}
    \label{fig:interviewBars}
\end{figure}

We asked the participants which password scheme they thought was the safest and the most difficult to guess or hack. 
 Figure~\ref{fig:interviewBars} shows a very similar pattern between children and adults, in terms of their perceived safety of the schemes, and no significant effect of scheme was found.

\subsubsection{Understanding of Security}
We found that compared to children, adults have a much better understanding of online privacy and security, and they reported well-known security advice, such as, not sharing passwords with others, frequently changing passwords, and avoiding logging in to untrusted websites. On the other hand, only a small number of children reported any advice. Among those who did, they explained that they should not share their passwords with anyone and should not give their personal information to strangers, they should have strong passwords stored in a safe place, and they should not click on ads or give their email addresses to strangers. The majority of children rely on their parents for managing their passwords; they either share their passwords with their parents, or they ask their parents to reset their passwords for them if they have trouble remembering them. Only six children manage their passwords on their own, and they reported writing them down. We found, contrary to previous research~\cite{Read:2012:DTP:2307096.2307125, doi:10.1080/07380569.2013.807719}, that the majority of children are fairly unaware of security practices and may be particularly vulnerable to attacks.

\subsubsection{Observations}

While using the adapted PassTiles interface, three children found it difficult to use the mouse and opted to use the touchscreen for inputting their passwords. All in all, children did not show any signs of confusion using the interface and were able to perform the tasks without major problems. The adults similarly did not encounter any significant hurdles. 

However, we observed that \emph{both} children and adults made the mistake of clicking the same tile twice during login, without noticing the number-of-clicks counter. This happened by either double-clicking a tile (rather than a single click) or by forgetting the tile was clicked, and clicking it again. Clicked tiles are not highlighted during the login phase to help protect users from shoulder surfing attacks. 

Additionally, the two youngest participants (7 years old) had trouble pronouncing the Words password, making it difficult for them to identify their tiles. On the other hand, several adult participants and the older child participants tried to memorize the Words and Objects password by forming a sentence with the words, or the objects' names.

Finally, we observed that children overlooked the pop-up message notifying them whether their login attempt was successful. This feedback, which was intended to help with the formation of a correct mental model of the system, was noticed by almost all adult participants but ignored by the majority of the children.


\section{Discussion}\label{sec:discussion}

The majority of children learned about passwords from their parents or older siblings. However, advice they receive from adults is not necessarily ideal; adults struggle with passwords and they do not always follow security best practices, which even applies to security experts~\cite{Stobert2016} and system administrators~\cite{Abdou2016}. Moreover, coping strategies that adults use are not necessarily applicable or useful for children. For example, to easily memorize the Words password, some of the adults formed a sentence with the words appearing on the highlighted tiles. However, we observed that children were struggling with this scheme; they took the longest logging in using Words and they had much trouble memorizing their Words password. Therefore, when designing authentication systems for children, we especially need to think about their skills, cognitive abilities, and preferences, since these can have significant impact on children's ability to use the system.

We found that, when choosing an image for Image PassTiles, most adults chose an image they thought would be most helpful in remembering---having a variety of content in different locations. On the other hand, children selected images they found to be interesting, usually containing animals or toys, and ignoring whether it would be easy to remember when divided into tiles.  Cued recall works best with unique cues on the background image, and since children are inattentive to this fact, it is important that the pool of images for ImagePassTiles should have enough distinct features to help with memorability. In addition, if we allow users to upload their own images, as suggested by some of the children participating in our study, these images should undergo a prescreening process to make sure they are feature-rich.

During the memorization phase, adults generally did not practice entering their passwords as often as children did; they spent the time observing the password carefully. Perhaps because adults are more familiar with the process of learning passwords than children, they may have been better able to draw on these past experiences to come up with successful memorization strategies. Adults perceived Image PassTiles to be harder to memorize, and so they spent significantly more time memorizing it than the other two schemes. On the other hand, children took their time to memorize their passwords and practiced them several times before logging in. Despite practicing, they had trouble remembering their passwords; we may need to offer additional support to guide their efforts. For instance, it might be possible to extend password memorization through gamification to achieve better success. 

Overall, the login success rates were much lower than would be necessary for a real world deployment, which could be due to the structure of our study. The scenario was more difficult than would be in a real life situation, \textit{e.g.,} users do not normally have to memorize three different passwords in a very short time frame. Thus, in the real world, we would expect the login success rates to increase. However, we can see this is an especially hard task for children. 
Based on our experiences and observations, we make the following recommendations for children's authentication.

\subsubsection*{R1. Facilitate memorization through learning and training features}
We recommend incorporating learning/training features to facilitate memorization within the password creation process. Due to children's inattentiveness to system notifications, and because interactive training features might prove to be more successful for children~\cite{Antle:2009:LIE:1487632.1487639}, we would recommend a more attractive method of notification that includes positive feedback from the system as encouragement. We would also like to encourage the creation of stories during the memorization process of both Objects and Words PassTiles by linking together the password's words or objects and incorporating this strategy into the user interface.

\subsubsection*{R2. Adapt interface to be age-appropriate for children}
We observed age-related differences in how children memorized their passwords and interacted with the interface. As with earlier research~\cite{bruckman2002hci}, our work indicates that we need different computer interfaces for children throughout the stages of development. For example, for younger children, we could use colourful words from age appropriate curriculum or from popular storybooks in Words PassTiles, and familiar objects such as animals and toys in Objects PassTiles. To help with their interaction with graphical passwords, we recommend using touchscreen input instead of mouse input as children tend to accidentally drag and double-click~\cite{bruckman2002hci} with the mouse which could lead to false input. Additionally, it might be useful, for children as well as adults, to refrain from counting duplicate clicks on the same tile towards the total count of inputed password tiles.

Although participants had less than desirable login success rates, looking more closely we see that most participants had nearly correct entries. This may indicate that shorter passwords might be more manageable for children (albeit at a decrease in security). Alternatively, a system which accepts entries as correct if they meet a certain threshold may also be conceivable, which aligns with Chatterjee~\textit{et al.}'s~\cite{chatterjee2016password} typo-tolerant password authentication framework.

\subsubsection*{R3. Combine schemes to improve memorability}
We were surprised that adults found Words PassTiles to be the most difficult scheme. We had assumed that because adults have a bigger vocabulary and are more experienced with words, they would find Words PassTiles more favourable. Although children generally had difficulty, the ones who formed stories with the words in their passwords remembered them better. We recommend using words with pictures to provide additional cues that could help with memorization. This could also be beneficial in circumstances where a user does not recognize the object illustrated.


\section{Conclusion}\label{sec:conclusion}

Despite the abundance of work looking at authentication mechanisms, the literature addressing children's authentication is relatively sparse. We explored graphical passwords as an alternative authentication method for children through two user studies---one with children and another with adults. Both children and adults preferred graphical passwords to their existing authentication scheme. Of the three schemes, children were most successful with Objects PassTiles, although they had some difficulty completely recalling any of their passwords. We explored similarities and differences in preferences and performance between children and adults. For example, we found that children are less careful when memorizing their passwords than adults and memorization strategies that adults use are not always applicable for children. Thus, it is necessary to consider children's cognitive abilities and skills to design child-friendly authentication systems.

\end{document}